\documentclass[%
 reprint,
superscriptaddress,
bibnotes,
 amsmath,amssymb,
 aps,
]{revtex4-1}

\usepackage{dcolumn}
\usepackage{bm}
\usepackage{mathrsfs}
\usepackage[dvipdfmx]{graphicx}
\usepackage{bm}
\usepackage{here}
\usepackage{amssymb,amsfonts,amsmath,amsthm,stmaryrd}
\usepackage{enumerate,float,indentfirst}
\usepackage{color}
\usepackage{simplewick}
\usepackage{tikz}

\theoremstyle{definition}

\theoremstyle{remark}

\allowdisplaybreaks

\def\nn{\nonumber}

\def\Tr{{\rm Tr}\,}

\definecolor{red}{rgb}{1,0,0}
\definecolor{orange}{rgb}{1,0.5,0}
\definecolor{violet}{rgb}{0.7,0,1}

\def\cre{\color{red}}

\def\cg{\color{green}}
\def\cb{\color{blue}}

\newcommand{\bel}[1]{\begin{equation}\label{#1}}
\newcommand{\bal}[1]{\begin{eqnarray}\label{#1}}
\newcommand{\be}{\begin{equation}}
\newcommand{\ee}{\end{equation}}

\begin{document}

\preprint{FIAN/TD-12/19, ITEP/TH-33/19, IITP/TH-19/19, MIPT-TH-17/19, NITEP 37, OCU-PHYS 512}

\title{  {\Large Correspondence between Feynman diagrams and operators \\
  in quantum field theory that emerges from tensor model}
	}

\author{N. Amburg}
\email{amburg@mccme.ru}
 \affiliation{%
A.I. Alikhanov  Institute for Theoretical and Experimental Physics of NRC “Kurchatov Institute”, B. Cheremushkinskaya, 25, Moscow, 117259, Russia
}
\affiliation{%
National Research University “Higher School of Economics” , Myasnitskaya Ul., 20, Moscow, 101000, Russia
}%
\author{H. Itoyama}
\email{itoyama@sci.osaka-cu.ac.jp}
 \affiliation{%
	Nambu Yoichiro Institute of Theoretical and Experimental Physics (NITEP)
}
 \affiliation{%
 	Department of Mathematics and Physics, Graduate School of Science\\
 	Osaka City University
 }
 \affiliation{%
	Osaka City University Advanced Mathematical Institute (OCAMI)\\
	3-3-138, Sugimoto, Sumiyoshi-ku, Osaka, 558-8585, Japan
}%
\author{A. Mironov}
\email{mironov@itep.ru}
\affiliation{%
I.E.Tamm Theory Department, Lebedev Physics Institute, Leninsky prospect, 53, Moscow 119991, Russia
}%
\affiliation{%
A.I. Alikhanov  Institute for Theoretical and Experimental Physics of NRC “Kurchatov Institute”, B. Cheremushkinskaya, 25, Moscow, 117259, Russia
}%
\affiliation{%
Institute for Information Transmission Problems of RAS (Kharkevich Institute), Bolshoy Karetny per. 19, build.1, Moscow 127051 Russia
}%
\affiliation{%
Kavli Institute for Theoretical Physics, Konh Hall, UC Santa Barbara, CA 93106-4030 USA
}%
\author{A. Morozov}
\email{morozov.itep@mail.ru}
\affiliation{%
Moscow Institute of Physics and Technology, Dolgoprudny, 141701, Russia
}%
\affiliation{%
A.I. Alikhanov  Institute for Theoretical and Experimental Physics of NRC “Kurchatov Institute”, B. Cheremushkinskaya, 25, Moscow, 117259, Russia
}%
\affiliation{%
Institute for Information Transmission Problems of RAS (Kharkevich Institute), Bolshoy Karetny per. 19, build.1, Moscow 127051 Russia
}%
\affiliation{%
Kavli Institute for Theoretical Physics, Konh Hall, UC Santa Barbara, CA 93106-4030 USA
}%
\author{D. Vasiliev}
\email{dmtrvass@gmail.com}
 \affiliation{%
A.I. Alikhanov  Institute for Theoretical and Experimental Physics of NRC “Kurchatov Institute”, B. Cheremushkinskaya, 25, Moscow, 117259, Russia
}%
\affiliation{%
Institute for Information Transmission Problems of RAS (Kharkevich Institute), Bolshoy Karetny per. 19, build.1, Moscow 127051 Russia
}%
\affiliation{%
Moscow Institute of Physics and Technology, Dolgoprudny, 141701, Russia
}%
\author{R. Yoshioka}%
 \email{yoshioka@sci.osaka-cu.ac.jp}
 \affiliation{%
	Osaka City University Advanced Mathematical Institute (OCAMI)\\
	3-3-138, Sugimoto, Sumiyoshi-ku, Osaka, 558-8585, Japan
}%



\date{\today}

\begin{abstract}
  A novel functorial relationship in perturbative quantum field theory is pointed out that associates Feynman diagrams (FD) having no external line in one theory ${\bf Th}_1$ with singlet operators in another one ${\bf Th}_2$ having an additional $U({\cal N})$ symmetry and is illustrated by the case where ${\bf Th}_1$ and ${\bf Th}_2$ are respectively
 the rank $r-1$ and the rank $r$ complex tensor model. The values of FD in ${\bf Th}_1$ agree with
  the large ${\cal N}$ limit of the Gaussian average of those operators in ${\bf Th}_2$.
  The recursive shift in rank by this {\it FD functor} converts numbers into vectors, then into matrices,
   and then into rank $3$ tensors ${\cdots}$.
 This {\it FD functor} can straightforwardly act on the $d$ dimensional tensorial quantum field theory (QFT)  counterparts as well.
  In the case of rank 2-rank 3 correspondence, it can be combined with the geometrical
   pictures of the dual of the original FD, namely,  equilateral triangulations (Grothendieck's {\it dessins d'enfant})
   to form a triality which may be regarded as a bulk-boundary correspondence.
\end{abstract}

\pacs{Valid PACS appear here}
\maketitle



 In modern quantum field theory by functional integrals or operator products (see  \cite{Peskin}, for example), Gaussian averages of a set of  operators invariant under given symmetry  can be a starting point of the perturbative consideration.
 In the case of four spacetime dimensions, the criterion of perturbative renormalizability to tame UV divergences
  selects  the list of such operators to consist of a few ones,
  and we can start full-fledged perturbation theory based on the action in accord with the list.
 Physical $S$-matrix can be extracted by resorting to the state/operator correspondence.

  Over the years, we have, however, occasionally seen cases in the study of (supersymmetric) low energy effective actions where we may even enumerate the whole set of such singlet operators and lift our considerations to all possible vacua and
    perturbation theory thereon. This phenomenon is known to be handled by the matrix model in general \cite{DV,CM0209,IM0211} and
     the structure of chiral ring \cite{CDSW0211} is responsible for the reduction in the degrees of freedom from four dimensional quantum fields to zero dimensional matrices.
  In this letter, we point out a novel functorial relationship that lies in
   Feynman diagrams (FD) having no external line in one theory ${\bf Th}_1$  and
   singlet operators in another one ${\bf Th}_2$   having an additional $U({\cal N})$ symmetry.
   We will refer to this as {\it FD functor} from now on.

  We will present the {\it FD functor} in the rank $r$ complex tensor model which has been studied in  \cite{IMM,IMMcj,MM,IYosh1903,KPT1808}.
 The symmetry group is $\displaystyle{\prod_{i=1}^r} U({N}_i)$ (as opposed to $U(N)$)
    and the rank $r$ implies $r$ different colorings in the diagrams representing the operators.
  The cases with lower $r$ are the vector model  for $r=1$, the rectangular complex matrix model for $r=2$,
   and the Aristotelian tensor model for $r=3$, and so on.
The {\it FD functor} relates  Feynman diagrams (FD) in ${\bf Th}_1 =\bm{{\cal T}}_{r-1} $ with singlet operators in
${\bf Th}_2 =\bm{{\cal T}}_r$.
Accordingly, the number of entries of the former coincides with
 that of the latter :
\begin{align}\label{r/r-1}
&\# ({\text{ Feynman diagrams in}}\ \bm{{\cal T}}_{r-1})  \cr
&= \# ({\text{singlet operators in}}\ \bm{{\cal T}}_r)
\end{align}
This equality holds at each level, i.e. {\bf the number of Feynman diagrams with $n$ propagators in
$\bm{{\cal T}}_{r-1}$ is equal
to the number of singlet operators with $2n$ vertices in $ \bm{{\cal T}}_{r}$.} Moreover, at  the large ${\cal N}$ limit, values of the Feynman diagrams coincide with the
Gaussian averages of the singlet operators.
The purpose of this letter is to demonstrate this operator/FD correspondence fully.

Restricting ourselves to rank 2 - rank 3 correspondence, we will discuss
that the above {\it FD functor} can be combined with the geometrical pictures of the dual of the original FD, namely, the equilateral triangulations (Grothendieck's {\it dessins d'enfant}) \cite{LZ2013} to form a triality.
In particular, we will relate directly any rank 3 tensor operator with a {\it dessin} and the map associated with it and this triality provides a higher-rank point of view for the two different approaches to non-critical string theory (two-dimensional system coupled to $2d$ gravity),
which are the continuum one based on the direct integration over $2d$ metric \cite{Polyakov} and the discretized one given by the matrix model \cite{Migdal}.

In Feynman diagram calculations of local quantum field theory in $d$ dimensions,
 expressions for the values take a (somewhat trivial) factorized form of the part containing tensorial structure associated with the internal symmetry and the well-known part of the spacetime propagator.
This is a consequence of the theorem \cite{CM1967} in cases where the symmetry of fields are reflected in the symmetry of $S$-matrix.
We will, in fact, see easily that the above {\it FD functor} can act on the $d$ dimensional QFT generalization of the tensor model as well and this uplift should hold in a more general setup as long as the theorem applies.

Let us now consider the {$\bm{\mathcal{T}}_r/\bm{\mathcal{T}}_{r-1}$ correspondence.
The level $n$ operators in the rank $r$ model are obtained by contracting all indices of $n$ tensors $M_{a_1a_2\cdots a_r}$ with those of $n$ complex conjugates $\bar{M}^{a_1a_2\cdots a_r}$:
\begin{equation}\label{operator}
K_{\vec{\sigma}^{(r)}} = \prod_{p=1}^n M_{a_1^p\cdots a_r^p} \bar{M}^{a_1^{\sigma_1(p)}\cdots a_r^{\sigma_r(p)}},
\end{equation}
where the contraction is specified by an $r$-ple of permutations $\vec{\sigma}^{(r)} = (\sigma_1,\cdots,\sigma_r) \in S_n^{\otimes r}$.
The $i$-th index runs over $1, \cdots, N_i$.

On the other hand,
any Feynman diagram (no external line) with $n$ propagators is given by the Wick contractions of a level $n$ operator.
Let us consider the Wick contractions of a rank $r-1$ operator designated $K^{(n)}_{\vec{\sigma}^{(r-1)}}$ in $\bm{\mathcal{T}}_{r-1}$.
The corresponding Feynman diagram is specified by a permutation $\sigma \in S_n$,
\begin{equation}
\mathbf{FD}_{\sigma} \left(K_{\vec{\sigma}^{(r-1)}}\right) = \prod_{p=1}^n W _{p,\sigma(p)} \left(K_{\vec{\sigma}^{(r-1)}}\right),
\end{equation}
where
\begin{equation}
 W_{p,q}(M^{(p)}\bar{M}_{(q)}) =
 \contraction{}{M}{{}_{a_1^p\cdots a_r^p}}{\bar{M}}
 M_{a_1\cdots a_r}^{(p)}\bar{M}^{b_1\cdots b_r}_{(q)} = \prod_{i=1}^r \delta_{a_i}^{b_i},
\end{equation}
is the Wick contractions of the $p$-th $M$ and the $q$-th $\bar{M}$ in $K_{\vec{\sigma}^{(r-1)}}$.

Now let us form an $r$-component vector  $(\vec{\sigma}^{(r-1)}, \sigma) \in S_n^{\otimes r}$.
Recall that the operator \eqref{operator} is labeled by an $r$-ple of permutations and,
therefore, there is always a rank $r$ operator labeled by  $(\vec{\sigma}^{(r-1)}, \sigma)$.

Conversely, for any rank $r$ operator, by interpreting the contraction of any one of $r$ indices as a symbol specifying the Wick contraction pair,
we can also obtain the corresponding rank $r-1$ Feynman diagram.

Hence the following correspondence is established,
\begin{equation}
  \mathbf{FD}_{\sigma} \left(K_{\vec{\sigma}^{(r-1)}}\right) ~~ \leftrightarrow ~~ K_{(\vec{\sigma}^{(r-1)}, \sigma)}.
\end{equation}
This correspondence is one to one  and we obtain the relation \eqref{r/r-1}.


Let $\mathrm{Op}^{(r)}$ be the set of all operators in the rank $r$ tensor model and $\mathrm{FD}^{(r-1)}$  be the set of all Feynman diagrams in the rank $r-1$ tensor model.
For any operator $K$,
we define $FD_{i}$ $(i=1,\cdots r)$ by $\mathrm{Op}^{(r)}  \ni K \mapsto FD_{i}(K) \in  \mathrm{FD}^{(r-1)}$, mapping $K$ to the corresponding Feynman diagram.
Here, the subscript $i$ means that the contraction by the $i$-th index $a_i$ is interpreted as the Wick contraction symbol.
One may pick any $i$ as that does not change the nature or category of our functor and
we may denote $N_i$ by $\mathcal{N}$.

Let us consider the Gaussian average $\langle K \rangle$ for an operator $K \in \text{Op}^{(r)}$ at level $n$.
 The leading term in $N_i$ is order $n$ and
 comes from the case where all of the Wick contractions of ${M}$ and $\bar{M}$
 are accompanied with the $a_i$ index loops:
\begin{equation}
\langle K \rangle = N_i^n \prod_{j \neq i}  N_j^{P_j} + \mathcal{O}(N_i^{n-1}).
\end{equation}
Here, the exponent $P_j$ is some integer less than or equal to $n$.
On the other hand,  $FD_i (K)$ has the following contribution to the Gaussian average of the rank $r-1$ operator,
\begin{equation}
 \text{value}\left\{FD_i (K)\right\} = \prod_{j \neq i} N_j^{P_j}.
\end{equation}
We obtain
\begin{equation}\label{FD/OP average}
  \text{value}\left\{FD_i (K)\right\} = \lim_{N_i \to \infty} \frac{1}{N_i^n} \langle K \rangle,
\end{equation}
demonstrating that the Op/FD correspondence is realized as the relation between the Gaussian averages as well.
For any rank $r-1$ operators $O^{(r-1)}$, there is a linear combination of the operators $O^{(r)}$ in the rank $r$ model such that
\begin{widetext}
\begin{equation}\label{average}
\left< O^{(r-1)}\right> = \sum_{\sigma} {\rm value}\Big\{{\bf FD}_{\sigma}(O^{(r-1)})\Big\}
\ \stackrel{\rm FD/Op}{=}
\sum_{ops} \text{value} \left\{  FD_i (O^{(r)}) \right\}
=\ \lim_{{N_i}\to \infty} \frac{1}{N_i^n} \sum_{ops}  \left<O^{(r)}\right>  .
\end{equation}
\end{widetext}

For example, one can check the $\bm{\mathcal{T}}_3/\bm{\mathcal{T}}_2$ correspondence
\begin{align}
 \left\langle( \Tr A\bar{A})^2 \right\rangle
 &= \text{value}\{FD_{\rm blue} (K_{\cb 2})\} +\text{value}\{FD_{\rm blue}(K_1^2) \} \nn\\
 &= \lim_{{N_b} \to \infty} \frac{1}{{N_b}^2}
 \langle K_{\cb 2} + K_1^2 \rangle \nn\\
 &= N_rN_g + N_r^2N_g^2.  \\
 \left\langle \Tr (A\bar A)^2\right\rangle
 &= \text{value}\{FD_{\rm blue} (K_{\cre 2})\} + \text{value}\{FD_{\rm blue} (K_{\cg 2})\}  \nn\\
 &= \lim_{{N_b} \to \infty} \frac{1}{{N_b}^2}
 \langle K_{\cre 2} + K_{\cg 2} \rangle \nn\\
  &= N_rN_g^2+N_r^2N_g.
\end{align}
See  \cite{IMMcj} for evaluating the averages.
Here the rank 2 tensors (rectangular matrices) are denoted by $A$ and $\bar{A}$ and
\begin{align}
&K_1 = M_{{\cre a_1}{\cg a_2}{\cb a_3}}\bar{M}^{{\cre a_1}{\cg a_2}{\cb a_3}} = ~
 \lower3ex\hbox{\begin{tikzpicture}
  \draw [thick,red] (1,0) arc (0:180:0.5) ;
  \draw [thick,green] (1,0) arc (0:-180:0.5) ;
  \draw[thick,blue](0,0)--(1,0);
  \filldraw [thick,fill=white] (0,0) circle (0.1);
  \filldraw [thick,fill=black] (1,0) circle (0.1);
  \end{tikzpicture}}~, \\
&K_{\cre 2} =
   \lower3ex\hbox{\begin{tikzpicture}
    \draw[thick,red](0,0.4)--(0.8,0.4);
    \draw[thick,red](0,-0.4)--(0.8,-0.4);
    \draw [thick,green] (0,0.4) to [out=180, in=180] (0,-0.4);
    \draw [thick,blue] (0,0.4) to [out=0, in=0] (0,-0.4);
    \draw [thick,blue] (0.8,0.4) to [out=180, in=180] (0.8,-0.4);
    \draw [thick,green] (0.8,0.4) to [out=0, in=0] (0.8,-0.4);
    \filldraw [thick,fill=white] (0,0.4) circle (0.1);
    \filldraw [thick,fill=black] (0.8,0.4) circle (0.1);
    \filldraw [thick,fill=black] (0,-0.4) circle (0.1);
    \filldraw [thick,fill=white] (0.8,-0.4) circle (0.1);
    \end{tikzpicture}}, ~
   K_{\cg 2} =
      \lower3ex\hbox{\begin{tikzpicture}
       \draw[thick,green](0,0.4)--(0.8,0.4);
       \draw[thick,green](0,-0.4)--(0.8,-0.4);
       \draw [thick,red] (0,0.4) to [out=180, in=180] (0,-0.4);
       \draw [thick,blue] (0,0.4) to [out=0, in=0] (0,-0.4);
       \draw [thick,blue] (0.8,0.4) to [out=180, in=180] (0.8,-0.4);
       \draw [thick,red] (0.8,0.4) to [out=0, in=0] (0.8,-0.4);
       \filldraw [thick,fill=white] (0,0.4) circle (0.1);
       \filldraw [thick,fill=black] (0.8,0.4) circle (0.1);
       \filldraw [thick,fill=black] (0,-0.4) circle (0.1);
       \filldraw [thick,fill=white] (0.8,-0.4) circle (0.1);
       \end{tikzpicture}}, ~
   K_{\cb 2} =
      \lower3ex\hbox{\begin{tikzpicture}
       \draw[thick,blue](0,0.4)--(0.8,0.4);
       \draw[thick,blue](0,-0.4)--(0.8,-0.4);
       \draw [thick,red] (0,0.4) to [out=180, in=180] (0,-0.4);
       \draw [thick,green] (0,0.4) to [out=0, in=0] (0,-0.4);
       \draw [thick,green] (0.8,0.4) to [out=180, in=180] (0.8,-0.4);
       \draw [thick,red] (0.8,0.4) to [out=0, in=0] (0.8,-0.4);
       \filldraw [thick,fill=white] (0,0.4) circle (0.1);
       \filldraw [thick,fill=black] (0.8,0.4) circle (0.1);
       \filldraw [thick,fill=black] (0,-0.4) circle (0.1);
       \filldraw [thick,fill=white] (0.8,-0.4) circle (0.1);
       \end{tikzpicture}}.
\end{align}

Let us specifically consider the $\bm{\mathcal{T}}_3/\bm{\mathcal{T}}_2$ correspondence.
We will see that the correspondence in this case extends to the triality including a relation with \textit{dessin}.
A \textit{dessin} $D$ is a graph embedded into a compact orientable surface $X$
 satisfying the following conditions:
\begin{itemize}
\item It is constructed by bicolored (red and green in this letter) vertices
 and edges.
\item Each edge connects the vertices which have the different color.
\item Multiple edges can end on a vertex.
\item The complement $X\backslash D$ is a disjoint union of the connected components
 which are called faces. Each face is homeomorphic to an open disk.
\end{itemize}

The \textit{dessins} are closely related to the so-called Belyi functions.
The Belyi function is a meromorphic function $\beta: X \to \mathbb{C}P^1$ unramified outside $\{0,1,\infty\}$. The pair $(\beta,X) $ is called Belyi map.
It is well-known that there is one to one correspondence between \textit{dessins} and Belyi maps up to  an automorphism of $X$.
For any \textit{dessin} $D$, there exists a Belyi function $\beta$ on $X$ and $D = \beta^{-1} ([0,1])$. the concrete relations are as follows:
\begin{itemize}
\item the red vertices are  $\beta^{-1}(0).$
\item  the green vertices  are $\beta^{-1}(1).$
\item the edges are $\beta^{-1} ([0,1])$
\end{itemize}
The number of edges ending on each vertex is equal to the ramification index and
there is only one pole on each face.

The operator/\textit{dessin} correspondence is as follows:
Every operator of level $n$ contains $n$ blue lines.
Let us choose one of the blue lines.
There are a red-blue cycle and a green-blue cycle
which share the blue line as the common boundary.
In particular,  these two faces can have the same orientation.
Therefore,
when we paint the regions surrounded by the red-blue (resp. green-blue) cycles
in red (resp. green) for any connected rank $3$ operator,
 the diagram representing the operator becomes an orientable surface
 painted in two colors.
The blue lines are the common boundaries of two adjacent regions.
In order to obtain the corresponding  \textit{dessin},  the following replacements should be made:
\begin{equation}\label{op-dessin}
\begin{tabular}{ccc}
\text{operator} && \textit{dessin} \\
red face&$\leftrightarrow$&red vertex \\
green face&$\leftrightarrow$&green vertex \\
blue line&$\leftrightarrow$&edge connecting bicolored vertices
\end{tabular}
\end{equation}
 For example,
\begin{equation} \label{K1}
 K_1= ~
 \lower3ex\hbox{\begin{tikzpicture}
  \draw [thick,red] (1,0) arc (0:180:0.5) ;
  \draw [thick,green] (1,0) arc (0:-180:0.5) ;
  \draw[thick,blue](0,0)--(1,0);
  \filldraw [thick,fill=white] (0,0) circle (0.1);
  \filldraw [thick,fill=black] (1,0) circle (0.1);
  \end{tikzpicture}}
  ~ = ~
 \lower3ex\hbox{\begin{tikzpicture}
 \filldraw [draw=red, fill=red] (1,0) arc (0:180:0.5) ;
 \filldraw [draw=green, fill=green] (1,0) arc (0:-180:0.5) ;
 \draw[ultra thick,blue](0,0)--(1,0);
 \end{tikzpicture}}
  ~~ \leftrightarrow ~~
  \lower3ex\hbox{\begin{tikzpicture}
   \draw[thick] (0,0.4) -- (0,-0.4);
   \filldraw [thick,fill=red] (0,0.4) circle (0.1);
   \filldraw [thick,fill=green] (0,-0.4) circle (0.1);
   \end{tikzpicture}}
  ~~ \leftrightarrow ~~
\beta (x) = x,
\end{equation}
where the function $\beta: {\mathbb{C}P}^1 \to {\mathbb{C}P}^1$ is the Belyi function associated to the \textit{dessin}.
Similarly,
\begin{alignat}{5}
  &K_{\cre 2} &&\leftrightarrow &\quad
   & \begin{tikzpicture}
     \draw[thick] (0,0) -- (1,0);
     \filldraw [thick,fill=green] (0,0) circle (0.1);
     \filldraw [thick,fill=red] (0.5,0) circle (0.1);
     \filldraw [thick,fill=green] (1,0) circle (0.1);
     \end{tikzpicture} &
       &\leftrightarrow &\quad
     &\beta(x) = x^2,
     \\
  &K_{\cg 2} && \leftrightarrow &
     &\begin{tikzpicture}
     \draw[thick] (0,0) -- (1,0);
     \filldraw [thick,fill=red] (0,0) circle (0.1);
     \filldraw [thick,fill=green] (0.5,0) circle (0.1);
     \filldraw [thick,fill=red] (1,0) circle (0.1);
     \end{tikzpicture} &
            &\leftrightarrow &\quad
    &{\footnotesize \beta(x) = 4x(1-x),}\\
  &K_{\cb 2} &&\leftrightarrow &
     &\lower3ex\hbox{\begin{tikzpicture}
     \draw [thick] (1,0) arc(0:180:0.5);
     \draw [thick] (1,0) arc(0:-180:0.5);
     \filldraw [thick,fill=red] (0,0) circle (0.1);
     \filldraw [thick,fill=green] (1,0) circle (0.1);
     \end{tikzpicture}   } &
            &\leftrightarrow &\quad
     &\beta(x) = \frac{(x+1)^2}{4x},
	\label{Krug2} \\
&K_{\cre 3} &&\leftrightarrow &
     &\lower3ex\hbox{\begin{tikzpicture}
     \draw[thick] (0,0) -- (0.5,0.3);
     \draw[thick] (0.5,0.3) -- (1,0);
     \draw[thick] (0.5,0.3) -- (0.5,1);
     \filldraw [thick,fill=green] (0,0) circle (0.1);
     \filldraw [thick,fill=red] (0.5,0.3) circle (0.1);
     \filldraw [thick,fill=green] (1,0) circle (0.1);
     \filldraw [thick,fill = green]  (0.5,1) circle (0.1);
     \end{tikzpicture}   } &
            &\leftrightarrow &\quad
      &\beta(x) = x^3,\\
  &K_{\cg 3} &&\leftrightarrow &
     &\lower3ex\hbox{\begin{tikzpicture}
     \draw[thick] (0,0) -- (0.5,0.3);
     \draw[thick] (0.5,0.3) -- (1,0);
     \draw[thick] (0.5,0.3) -- (0.5,1);
     \filldraw [thick,fill=red] (0,0) circle (0.1);
     \filldraw [thick,fill=green] (0.5,0.3) circle (0.1);
     \filldraw [thick,fill=red] (1,0) circle (0.1);
     \filldraw [thick,fill = red]  (0.5,1) circle (0.1);
     \end{tikzpicture}   } &
            &\leftrightarrow &\quad
      &\beta(x) = x^3+1,
    \\
 &K_{\cb 3} &&\leftrightarrow &
     &\lower3ex\hbox{\begin{tikzpicture}
     \draw [thick] (1,0) arc(0:180:0.5);
     \draw [thick] (1,0) arc(0:-180:0.5);
     \draw [thick] (0,0)--(1,0);
     \filldraw [thick,fill=red] (0,0) circle (0.1);
     \filldraw [thick,fill=green] (1,0) circle (0.1);
     \end{tikzpicture}   }&
            &\leftrightarrow &\quad
     &\beta(x) = \frac{(x+1)^3}{2(3x^2+1)},
\\
 &K_{{\cre 2} {\cg 2}} &&\leftrightarrow &
    &\begin{tikzpicture}
     \draw[thick] (0,0) -- (1.5,0);
     \filldraw [thick,fill=red] (0,0) circle (0.1);
     \filldraw [thick,fill=green] (0.5,0) circle (0.1);
     \filldraw [thick,fill=red] (1,0) circle (0.1);
    \filldraw [thick,fill=green] (1.5,0) circle (0.1);
     \end{tikzpicture}&
            &\leftrightarrow &\quad
      &\beta(x) = \frac{4x^3-3x+1}{2},\\
  &K_{{\cg 2} {\cb 2}} &&\leftrightarrow &
     &\lower1ex\hbox{\begin{tikzpicture}
     \draw [thick] (0,0) to [out=90, in=90] (0.7,0);
     \draw [thick] (0,0) to [out=-90, in=-90] (0.7,0);
     \draw[thick] (0.7,0)--(1.2,0);
     \filldraw [thick,fill=red] (0,0) circle (0.1);
     \filldraw [thick,fill=green] (0.7,0) circle (0.1);
     \filldraw [thick,fill=red] (1.2,0) circle (0.1);
     \end{tikzpicture}   } &
            &\leftrightarrow &\quad
       &\beta(x) = -\frac{(x+1)^2(x-8)}{27x},    \\
  &K_{{\cb 2} {\cre 2}} &&\leftrightarrow &
     &\lower1ex\hbox{\begin{tikzpicture}
     \draw [thick] (0,0) to [out=90, in=90] (0.7,0);
     \draw [thick] (0,0) to [out=-90, in=-90] (0.7,0);
     \draw[thick] (0.7,0)--(1.2,0);
     \filldraw [thick,fill=green] (0,0) circle (0.1);
     \filldraw [thick,fill=red] (0.7,0) circle (0.1);
     \filldraw [thick,fill=green] (1.2,0) circle (0.1);
     \end{tikzpicture}   } &
            &\leftrightarrow &\quad
      &\beta(x) = \frac{4(x-1)^3}{27x},
      \\
  &K_{3W} &&\leftrightarrow &
    & \lower3ex\hbox{\begin{tikzpicture}
     \draw [thick] (1,0) arc(0:180:0.5);
     \draw [thick] (0,0) to [out=0,in=120] (0.5,-0.3) to [out=-60,in=-180] (0.8,-0.5) to
     [out=0,in=-90](1,0);
     \draw [thick] (0,0) to [out=-90,in=-180] (0.2,-0.5) to [out=0,in=-120](0.45,-0.3);
     \draw [thick] (0.5,-0.2) to [out=60,in=-180] (1,0);
     \filldraw [thick,fill=red] (0,0) circle (0.1);
     \filldraw [thick,fill=green] (1,0) circle (0.1);
     \end{tikzpicture}   } &&
\end{alignat}
The \textit{dessin} corresponding to $K_{3W}$ is non-planar and
 can be embedded in a two-torus  as shown in Figure \ref{torus}.
The Belyi function is given by $\beta = \frac{1}{2} (1+y)$ on $X$: $y^2 = x^3+1$, $x,y \in \mathbb{C}$ \cite{AADKK0710}.
 \begin{figure}[H]
  \centering
  \includegraphics[height=2cm]{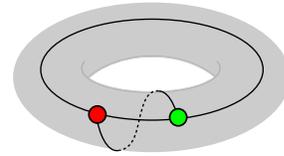}
  \caption{the \textit{dessin} embedded in a two-torus}\label{torus}
 \end{figure}%
We have checked the correspondence up to level 4.
The numbers of the connected operators at each level are $1,3,7,26,\cdots$.
Here 6 of 26 operators at level 4 correspond to \textit{dessins} on torus.
The operators of genus $g \geq 2$ appear at level 5 or higher.
These are, of course, in accord with the Riemann-Hurwitz formula
\begin{equation}
2g-2 = n - { V_R} - { V_G} - { F},
\end{equation}
where $n$ is the number of edges (the level of the operator),
$V_R$ and $V_G$ are the numbers of red and green vertices respectively and
 $F$ is the number of faces \cite{JRRG1012} .

The \textit{dessins} and the rank two Feynman diagrams are dual to each other, namely,
\begin{equation}\label{dessin-FD}
\begin{tabular}{ccc}
\textit{dessin} && \text{Feynman diagram} \\
center of face &$\leftrightarrow$ & vertex \\
edge&$\leftrightarrow$&propagator \\
red vertex&$\leftrightarrow$&red face \\
green vertex&$\leftrightarrow$&green face
\end{tabular}
\end{equation}
The centers of faces are the poles of the  Belyi function  and each face contains only one pole.

For example,
\begin{equation}
 (K_1 ~ \leftrightarrow )  ~~
 \lower3ex\hbox{\begin{tikzpicture}
    \draw[thick] (0,0.4) -- (0,-0.4);
    \filldraw [thick,fill=red] (0,0.4) circle (0.1);
    \filldraw [thick,fill=green] (0,-0.4) circle (0.1);
    \end{tikzpicture}} ~=~
\lower4ex\hbox{\begin{tikzpicture}
    \draw[thick] (0,0.4) -- (0,-0.4);
    \filldraw [thick,fill=red] (0,0.4) circle (0.1);
    \filldraw [thick,fill=green] (0,-0.4) circle (0.1);
    \draw[red, thick] (-0.3,0.1) rectangle (0.3,0.7);
    \draw [-latex,red] (0.2,0.1) -- +(0:0.1);
    \draw[green, thick] (-0.3,-0.1) rectangle (0.3,-0.7);
    \draw [-latex,green] (0.2,-0.1) -- +(180:0.1);
    \node at (0.5,0) {$\otimes$} ;
    \end{tikzpicture}}
  ~ \leftrightarrow ~
\lower3ex\hbox{\begin{tikzpicture}
    \draw [red, thick] (0,0) circle (0.3);
    \draw [green, thick] (0,0) circle (0.5);
    \draw [-latex,red] (0,-0.3) -- +(0:0.1);
    \draw [-latex,green] (0,-0.5) -- +(180:0.1);
    \node at (0.4,0) {$\otimes$} ;
    \end{tikzpicture}}
    = \Tr
     \contraction{}{A}{}{\bar{A}}
     A\bar{A}  = {\cre N_1}{\cg N_2},
\end{equation}
where the center of the face is denoted by $\otimes$.
This \textit{dessin} is embedded in a sphere and there exists, therefore, one face.
Note that the exponents of ${\cre N_1}$ and ${\cg N_2}$ are
 equal to the numbers of red and green vertices respectively.
This is because the vertices in a \textit{dessin} correspond to the faces in the corresponding Feynman diagram.

The above discussion provides the following bulk-boundary correspondence at the range (``the target space") of the Belyi function and its uplift to the rank 3 operators.
The vertices of each FD lie at infinity, namely, at the ``boundary" of the complex plane.
As for the propagators, they start and end at infinity where the vertices lie.
They must also cross the segment $[0,1]$ as they are dual to the edges, and, therefore, extends over the entire complex plane.
See \cite{Gopa1104} for a similar discussion.

Finally, it is straightforward to extend the Op/FD correspondence to the $d$ dimensional   field theory
described by the spacetime dependent tensors $M_{a_1a_2\cdots}(x)$ and $\bar{M}^{a_1a_2\cdots}(x)$ .
Let us consider the large $\mathcal{N}$ limit of the average of a singlet operator in the rank $r$ theory:
\begin{equation}\label{op-tft}
 \lim_{\mathcal{N} \to \infty} \frac{1}{\mathcal{N}^n} \int dx \langle O_{r,n}(x) \rangle,
\end{equation}
where $O_{r,n}(x)$ is a level $n$ operator.
Following the discussion leading to \eqref{average},
it is easy to see that \eqref{op-tft} has the corresponding Feynman diagram in the rank $r-1$ $d$ dimensional QFT.
The average \eqref{op-tft} of level $n$ operator includes $n$ propagators, each of which contains a momentum dependent factor.
As discussed above, the tensorial structure determines the corresponding Feynman diagram which has no external line and, therefore, all momenta have to be integrated.

For example, the level 3 operator $K_{3W}(x)$ in the rank 3 theory corresponds to a non planar graph:
\begin{align}
 \includegraphics[height=2cm]{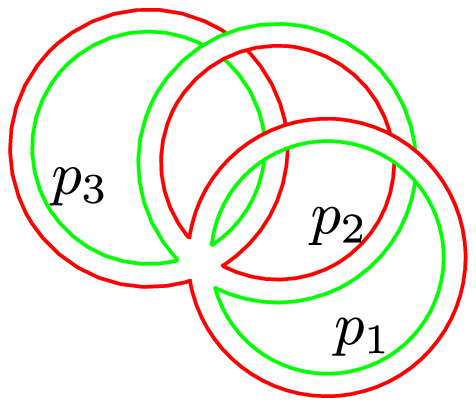} &= \int dp_1dp_2dp_3 G(p_1)G(p_2)G(p_3) N_rN_g \cr
 &= \lim_{{N_b}\to \infty}  \frac{1}{V N_b^3} \int dx \langle K_{3W}(x) \rangle .
\end{align}
Here $G(p)$ is the momentum  part of the propagator.

\begin{acknowledgments}
A. Mironov is grateful for the hospitality of NITEP, Osaka City University as well as that of the Workshop
New Trends in Integrable Systems 2019 held there during the period of September, 9-20. Our work is partly
supported by JSPS KAKENHI grant Number 19K03828 (H.I.) and OCAMI MEXT Joint Usage/Research Center on
Mathematics and Theoretical Physics (H.I., R.Y.), by the grant of the Foundation for the Advancement of Theoretical
Physics ``BASIS" (A.Mir., A.Mor.), by RFBR and NSFB according
to the research project 19-51-18006 (A.Mir., A.Mor.),
by RFBR grants 19-02-00815 (N.A., A.Mor.), 19-01-00680 (A.Mir.), RFBR 18-02-01081 (D.V.), by
joint grants 19-51-53014-GFEN (A.Mir., A.Mor.), 19-51-50008-YaF (A.Mir.), 18-51-05015-Arm (A.Mir.,
A.Mor.), 18-51-45010-IND (A.Mir., A.Mor.). A. Mironov and A. Morozov also acknowledge the hospitality of KITP and partial support by the National Science Foundation under Grant No. NSF PHY-1748958 at the final stage of this project.
\end{acknowledgments}


\begin{thebibliography}{20}%
\bibitem{Peskin}
M. E. Peskin and D. V. Schroeder, {\it An introduction to
  quantum field theory}, (Westview Press, 1995).

\bibitem{DV}
R. Dijkgraaf and C. Vafa, Nucl. Phys. B {\bf 644}, 3 (2002), arXiv:hep-th/0206255;
{\bf 644}, 21 (2002), arXiv:hep-th/0207106.

\bibitem{CM0209}
L. Chekhov and A. Mironov,
Phys. Lett. B {\bf 552}, 293 (2003), arXiv:hep-th/0209085.

\bibitem{IM0211}
H. Itoyama and A. Morozov,
Nucl. Phys. B {\bf 657}, 53 (2003), arXiv:hep-th/0211245.

\bibitem{CDSW0211}
F. Cachazo, M. R. Douglas, N. Seiberg, and E. Witten,
JHEP {\bf 12}, 071 (2002), arXiv:hep-th/0211170.

\bibitem{IMM}
H. Itoyama, A. Mironov, and A. Morozov, Phys. Lett.
B {\bf 771}, 180 (2017), arXiv:1703.04983 [hep-th];
 JHEP {\bf 06}, 115 (2017), arXiv:1704.08648 [hep-th];
 arXiv:1910.03261 [hep-th].

\bibitem{IMMcj} H. Itoyama, A. Mironov, and A. Morozov, Nucl. Phys. B {\bf 932}, 52 (2018), arXiv:1710.10027 [hep-th].

\bibitem{MM}
A. Mironov and A. Morozov,
Phys. Lett. B {\bf 771}, 503 (2017), arXiv:1705.00976 [hep-th];
{\bf 774}, 210 (2017), arXiv:1706.03667 [hep-th].

\bibitem{IYosh1903}
 H. Itoyama and R. Yoshioka, Nucl. Phys. B {\bf 945}, 114681
(2019), arXiv:1903.10276 [hep-th].

\bibitem{KPT1808}
 I. R. Klebanov, F. Popov, and G. Tarnopolsky,
PoS {\bf TASI2017}, 004 (2018), arXiv:1808.09434 [hep-th]

\bibitem{LZ2013}
S. K. Lando and A. K. Zvonkin, {\it Graphs on surfaces and
their applications}, Vol. 141 (Springer Science \& Business
Media, 2013).

\bibitem{Polyakov}
 A. M. Polyakov, Phys. Lett. {\bf 103B}, 207 (1981).

\bibitem{Migdal}
A. A. Migdal, Phys. Rept. {\bf 102}, 199 (1983).

\bibitem{CM1967}
S. Coleman and J. Mandula, Phys. Rev. {\bf 159}, 1251
(1967).

\bibitem{AADKK0710}
 N. M. Adrianov, N. Ya. Amburg, V. A. Dremov, Yu. Yu.
Kochetkov, E. M. Kreines, Yu. A. Levitskaya, V. F. Nasretdinova,
and G. B. Shabat, J. Math. Sci. {\bf 158}, 22
(2009), arXiv:0710.2658 [math.AG].

\bibitem{JRRG1012}
V. Jejjala, S. Ramgoolam, and D. Rodriguez-Gomez,
JHEP {\bf 03}, 065 (2011), arXiv:1012.2351 [hep-th].

\bibitem{Gopa1104}
R. Gopakumar, (2011), arXiv:1104.2386 [hep-th].

\end{thebibliography}

%

\end{document}